\documentclass[sigplan,nonacm]{acmart}

\AtBeginDocument{%
  }

\usepackage{microtype}
\usepackage{graphicx}
\usepackage{booktabs} 

\usepackage{hyperref}

\usepackage[capitalize,noabbrev]{cleveref}

\theoremstyle{plain}

\theoremstyle{definition}

\theoremstyle{remark}

\usepackage[textsize=tiny]{todonotes}

\usepackage{float}
\usepackage{subfig}
\usepackage{comment}

\usepackage{xurl}

\usepackage{amsmath,amssymb,amsfonts}
\usepackage{algorithm}
\usepackage{algpseudocode} 

\algrenewcommand\algorithmicrequire{\textbf{Input:}}
\algrenewcommand\algorithmicensure{\textbf{Output:}}

\usepackage{tabularx}
\newcommand{\multiline}[1]{%
  \begin{tabularx}{\dimexpr\linewidth-\ALG@thistlm}[t]{@{}X@{}}
    #1
  \end{tabularx}
}

\newcommand{\circled}[1]{\tikz[baseline=(myanchor.base)] \node[circle,fill=.,inner sep=0.3pt] (myanchor) {\color{-.}\bfseries\footnotesize #1};}

\newcommand{\secref}[1]{\S\ref{#1}} 

\newcommand{\needupdate}[1]{\textcolor{black}{#1}}
\newcommand{\ours}{\textit{Adrenaline}}

\begin{document}

\title{Injecting \textit{Adrenaline} into LLM Serving: Boosting Resource Utilization and Throughput via Attention Disaggregation}

\author{Yunkai Liang}
\authornote{Both authors contributed equally to this research.}
\authornote{Work done during his internship at Huawei Cloud.}
\affiliation{%
  \institution{Sun Yat-sen University}
  \country{} 
}

\author{Zhangyu Chen}
\authornotemark[1]
\affiliation{%
  \institution{Huawei Cloud}
  \country{} 
}

\author{Pengfei Zuo}
\authornote{Corresponding authors are Pengfei Zuo (pengfei.zuo@huawei.com) and Zhi Zhou (zhouzhi9@mail.sysu.edu.cn).}
\affiliation{%
  \institution{Huawei Cloud}
  \country{} 
}

\author{Zhi Zhou}
\authornotemark[3]
\affiliation{%
  \institution{Sun Yat-sen University}
  \country{} 
}

\author{Xu Chen}
\affiliation{%
  \institution{Sun Yat-sen University}
  \country{} 
}

\author{Zhou Yu}
\affiliation{%
  \institution{Huawei Cloud}
  \country{} 
}
\begin{abstract}
In large language model (LLM) serving systems, executing each request consists of two phases: the compute-intensive prefill phase and the memory-intensive decoding phase. To prevent performance interference between the two phases, current LLM serving systems typically adopt prefill-decoding disaggregation, where the two phases are split across separate machines. However, we observe this approach leads to significant resource underutilization. Specifically, prefill instances that are compute-intensive suffer from low memory utilization, while decoding instances that are memory-intensive experience low compute utilization.
To address this problem, this paper proposes \textit{Adrenaline}, an attention disaggregation and offloading mechanism designed to enhance resource utilization and performance in LLM serving systems. \textit{Adrenaline}'s key innovation lies in disaggregating part of the attention computation in the decoding phase and offloading them to prefill instances. The memory-bound nature of decoding-phase attention computation inherently enables an effective offloading strategy, yielding two complementary advantages: (1) improved memory capacity and bandwidth utilization in prefill instances, and (2) increased decoding batch sizes that enhance compute utilization in decoding instances — collectively boosting overall system performance. \textit{Adrenaline} achieves these gains through three key techniques: low-latency decoding synchronization, resource-efficient prefill colocation, and load-aware offloading scheduling. Experimental results show that \textit{Adrenaline} achieves 2.28$\times$ higher memory capacity and 2.07$\times$ better memory bandwidth utilization in prefill instances, up to 1.67$\times$ improvements in compute utilization for decoding instances, and 1.68$\times$ higher overall inference throughput compared to state-of-the-art systems.
\end{abstract}

\maketitle

\section{Introduction}
\label{sec:introduction}

With the advanced transformer architectures and the development of billion-scale parameters, large language models (LLMs) have achieved remarkable success in content generation, with widespread applications in areas such as chatbots~\cite{DBLP:conf/nips/ZhengC00WZL0LXZ23}, writing assistant~\cite{DBLP:conf/sigir/SchaikP24}, and code generation~\cite{DBLP:conf/nips/LiuXW023}. Given that high-quality LLM inference typically requires high-performance GPU accelerators, many popular LLMs such as, GPT~\cite{DBLP:conf/nips/BrownMRSKDNSSAA20}, LLaMA \cite{DBLP:journals/corr/abs-2407-21783}, Qwen~\cite{DBLP:journals/corr/abs-2407-10671}, and Mistral~\cite{DBLP:journals/corr/abs-2401-04088}, are offered as cloud-based services to provide fast and accurate responses to inference queries. In these LLM serving systems, meeting
service level objectives (SLOs) for latency is critical to ensure that user requests are processed within strict deadlines. At the same time, providers seek to reduce serving costs by optimizing GPU utilization and maximizing throughput to serve as many requests as possible.



Executing each request in LLM serving systems involves two successive phases. The \emph{prefill phase}
computes all prompt tokens in parallel to generate the first
output token. The \emph{decoding phase} then sequentially predicts one output
token per model forward step based on the previous context, which includes the computed
intermediate results of the prompt and previously generated tokens, namely 
\emph{KV cache}. Due to the parallel processing of many tokens,
the prefill phase is typically compute-intensive, with latency measured by the time to the first token (TTFT). In contrast, the decoding
phase is memory-intensive due to the frequent loading of the KV cache with growing size, and its latency is measured by the time per output token (TPOT). Since a prefill step generally incurs much higher latency than a decoding step, running both phases on the same GPUs leads to significant interference, increasing TTFT for the prefill phase and TPOT for the decoding phase within the same batch~\cite{DBLP:conf/osdi/ZhongLCHZL0024}. 

To reduce interference between the prefill and decoding phases, modern LLM serving systems typically employ prefill-decoding (PD) disaggregation~\cite{DBLP:conf/isca/PatelCZSGMB24,
DBLP:conf/osdi/ZhongLCHZL0024,DBLP:journals/corr/abs-2407-00079}. By assigning prefill and decoding phases
to separate GPUs, PD disaggregation eliminates the interference between the two phases,
allowing each phase to meet its respective service-level objectives (SLOs) independently. 
Furthermore, disaggregating prefill
and decoding phases in different GPU pools enables automatic and
flexible resource scaling, accommodating the distinct resource requirements of each phase. Today, PD disaggregation is becoming the \emph{de-facto} practice of production LLM serving systems, such as vLLM~\cite{DBLP:conf/sosp/KwonLZ0ZY0ZS23}, NVIDIA Dynamo~\cite{NVIDIADynamo}, Kimi's Mooncake~\cite{DBLP:journals/corr/abs-2407-00079}, and Deepseek~\cite{liu2024deepseek}. 

\begin{figure}[t!]
	\centering
	\subfloat[\label{fig:prefillUtil} Prefill phase]{
		\includegraphics[width=0.46\linewidth]{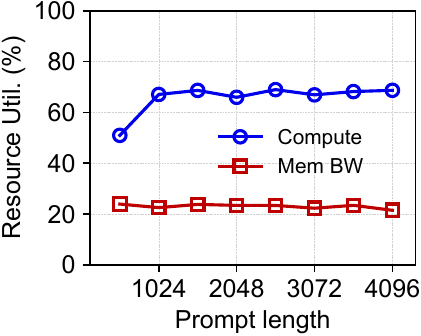}}\hfill
        \subfloat[\label{fig:decodeUtil} Decoding phase]{
            \includegraphics[width=0.46\linewidth]{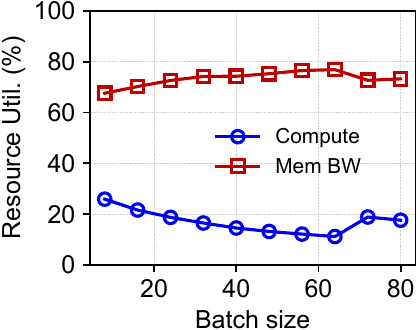}}
	\caption{\label{fig:PDUtil} \needupdate{Resource utilization in 
        disaggregated prefill and decoding phases for Llama-2 7B.
        The sequence length for the decoding phase is 1K.}}
\end{figure}

\begin{figure}[t!]
   \begin{minipage}{0.46\linewidth}
     \centering
      \includegraphics[width=0.98\linewidth]{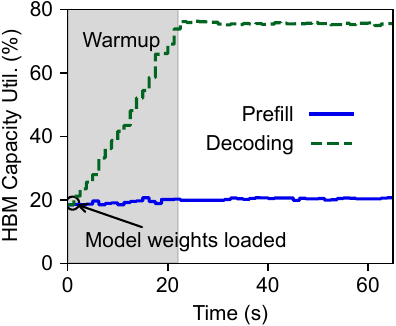}
      \vspace{-0.8em}
      \caption{\label{fig:PDHbmUtil} \needupdate{HBM utilization when 
        serving Llama-2 7B}.}
   \end{minipage}\hfill
   \begin{minipage}{0.48\linewidth}
     \centering
     \includegraphics[width=.98\linewidth]{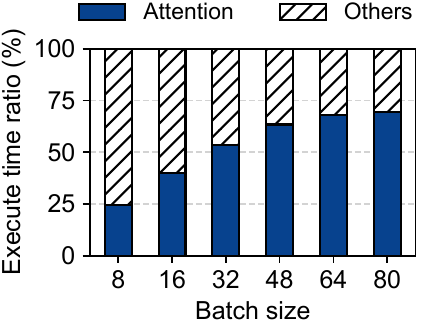}
     \vspace{-0.8em}
     \caption{\label{fig:AttnTimeRatio} \needupdate{The execution time of decoding attention}.}
   \end{minipage}
\end{figure}

However, we observe that PD disaggregation in LLM serving systems incurs significant underutilization of GPU resources, either in terms of compute or memory. Specifically, GPUs running the compute-intensive prefill phase often experience low utilization of their HBM capacity and bandwidth. In contrast, the memory-intensive decoding phase struggles with low compute resource utilization. As shown in Figure~\ref{fig:prefillUtil}, HBM bandwidth utilization of the prefill instance is below 30\%. In addition, Figure~\ref{fig:PDHbmUtil} shows that HBM capacity utilization is under 21\% in the prefill instance of existing PD disaggregation system, i.e.,  vLLM~\cite{DBLP:conf/sosp/KwonLZ0ZY0ZS23}. Meanwhile, compute utilization in the decoding instance is below 26\% for different batch sizes. Given the high cost of GPUs, such low resource utilization leads to increased inference costs. To investigate the causes of this underutilization, we measured the resource consumption of internal components in both the prefill and decoding phases. Our experimental results (\secref{sec:motivation}) reveal that four main components including QKV projection, attention computation, output projection, and feed-forward network in the prefill phase are all compute-intensive. In the decoding phase, Figure~\ref{fig:PDHbmUtil} shows HBM capacity utilization ratio is 75.5\% after warmup (\texttt{gpu\_memory\_utilization} is 0.8) and the KV cache accounts for 57.3\% HBM capacity for attention computation. Moreover, the memory-intensive attention kernel consumes lots of GPU time (e.g., 69.5\% of execution time per transformer layer when the batch size is 80 in Figure~\ref{fig:AttnTimeRatio}). In general, the high memory resource consumption in the decoding instance is mainly due to the storage and access of the KV cache for attention computation~\cite{DBLP:conf/usenix/GaoHSKJDYYZ24}.

Based on the observations above, this paper proposes
\ours\footnote{Adrenaline is a hormone that rapidly enhances the body's responsiveness. Similarly, our proposed technique acts like adrenaline injected into LLM serving systems to boost inference throughput.}, an \textbf{A}ttention \textbf{d}isagg\textbf{re}gatio\textbf{n} \textbf{a}nd off\textbf{l}oad\textbf{in}g m\textbf{e}chanism for LLM serving systems with PD disaggregation, to efficiently improve the GPU resource utilization and thus enhance the LLM
inference performance. 
\ours{} disaggregates and offloads part of attention computation in the decoding phase to GPUs running the prefill phase. 
This approach enhances the utilization of HBM capacity and bandwidth in the prefill instances by offloading the memory-intensive attention computation tasks from the decoding phase. Moreover, it increases the total batch size (the sum of the local batch size and the offloaded batch size) in the decoding instances, thereby improving their compute utilization.
However, several challenges arise in achieving attention disaggregation and offloading.

\textbf{\textit{1) Additional attention synchronization overheads in the decoding phase.}}
Since only part of the decoding attention tasks are offloaded,
the remaining decoding attention tasks are computed in the original
decode phase. As a result, the offloaded attention must be
synchronized with the rest of the decoding computation within a single decoding attention step (typically less than $1~ms$).
Any delay in this synchronization can lead to \emph{attention stall} in each transformer layer,
thereby increasing the TPOT.

\textbf{\textit{2) Performance interference in prefill instances.}}
Offloaded decoding attention tasks share GPU resources with
the prefill phase. For example, offloaded KV caches occupy HBM capacity, and
executing attention computation consumes both compute resources
and HBM bandwidth. The colocation of these tasks causes contention for GPU resources, which can
result in performance interference and reduced overall system efficiency.

\textbf{\textit{3) Efficient control of offloading rate.}}
Offloading too many attention computation tasks to the prefill instances can lead to excessive consumption of GPU
resources, which probably increases the TTFT. Conversely, offloading too few tasks may not justify the synchronization overhead.
Given the dynamic nature of sequence lengths~\cite{DBLP:conf/sosp/KwonLZ0ZY0ZS23} and the fluctuating load in prefill and decoding instances, the offloading strategy needs to be flexible. It should efficiently orchestrate the disaggregated LLM serving system to improve performance under real-world workloads.

To address these challenges, \ours{} incorporates three key techniques.
First, \ours{} utilizes a low-latency synchronization strategy by shortening the critical path of synchronizing the decoding phase and offloaded attention computation.  CUDA graph technologies are leveraged to reduce the extra kernel-launching overheads for attention offloading. 
Second, to mitigate performance interference, \ours{} explores compute-constrained attention computation and introduces a resource-efficient prefill colocation mechanism to isolate GPU resources. 
Finally, to determine the optimal offloading rate, \ours{} leverages a load-aware offloading scheduling strategy that dynamically adjusts the offloading rate. This adjustment is based on real-time compute and memory resource utilization in both the prefill and decoding phases, as well as inference SLOs. The offloading rate is continuously updated to respond to load fluctuations.

We have implemented \ours{} on top of
vLLM~\cite{DBLP:conf/sosp/KwonLZ0ZY0ZS23}.
Experimental results show that attention offloading of \ours{} achieves up to 2.07$\times$ memory utilization of prefill instances and 1.67$\times$ compute utilization in decoding instances for real-world workload (ShareGPT~\cite{sharegpt}). The overall output token throughput is enhanced by up to 1.68$\times$ compared to vLLM. The source code of \ours{} is released at Github\footnote{\url{https://github.com/ASISys/Adrenaline}}.
In summary, this paper makes the following contributions.
\begin{itemize}
  \item We identify and analyze the GPU resource underutilization problem in LLM serving systems with PD disaggregation and explore its underlying causes. 
  \item We propose \ours{}, a simple yet effective solution that enhances resource utilization and throughput in LLM serving systems through attention disaggregation and offloading.
  \item We introduce three key techniques in \ours{} including low-latency decoding synchronization, resource-efficient prefill colocation, and load-aware offloading scheduling to achieve efficient attention offloading.
  \item We implement and evaluate \ours{} on vLLM, demonstrating its effectiveness in improving performance and resource utilization.
\end{itemize}

\section{Background and Motivation}
\label{sec:background}

\subsection{Large Language Model Inference}
\label{sec:LLMBackground}

\textbf{Transformer Architecture.} The transformer network~\cite{DBLP:conf/nips/VaswaniSPUJGKP17} is a fundamental architecture in LLMs and has been
widely adopted in many popular models~\cite{DBLP:journals/corr/abs-2407-21783,DBLP:journals/corr/abs-2302-13971,DBLP:journals/corr/abs-2307-09288}.
The transformer consists of multiple repeated
transformer layers, each comprising four main computation steps:
(1) linear projection for the query ($q$), key ($k$), and value ($v$), collectively referred to as $QKV$ projection; 
(2) attention computation, also known as self-attention, using the KV caches; (3) linear projection of the attention output, called output projection;
(4) the feed-forward network (FFN). The KV caches store intermediate states during inference, i.e., computed $K$ and $V$, to avoid recomputing historical states for ongoing requests.

\textbf{Prefill and Decoding Phases.} LLM inference proceeds through two sequential phases: the prefill and decoding phases. Given a user prompt as input, the prefill phase processes the entire prompt to generate the first output token and populate the KV cache. The decoding phase then iteratively
generates subsequent output tokens by using the last output token as the input. The accumulated KV cache allows the decoding phase to compute only the $K$ and $V$ for the new token, enhancing efficiency.
The prefill phase is compute-intensive due to the parallel processing of multiple prompt tokens, and its latency is measured by the time to the first token (TTFT). In contrast, the decoding phase is memory-intensive, primarily due to the frequent loading of model weights and the growing size of the KV cache. Its latency is measured by the time per output token (TPOT).

\subsection{Prefill and Decoding Disaggregation}
\label{sec:DisaggBackground}

\begin{figure}[t!]
	\centering
	\subfloat[\label{fig:disaggOrigArch}PD Disaggregation]{
		\includegraphics[width=0.47\linewidth]{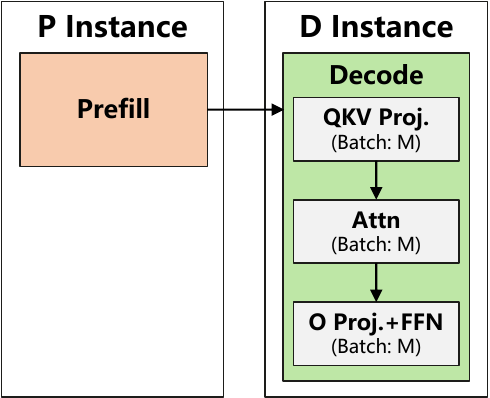}}
	\hfill
    \subfloat[\label{fig:disaggOffloadArch} \ours{}]{
		\includegraphics[width=0.47\linewidth]{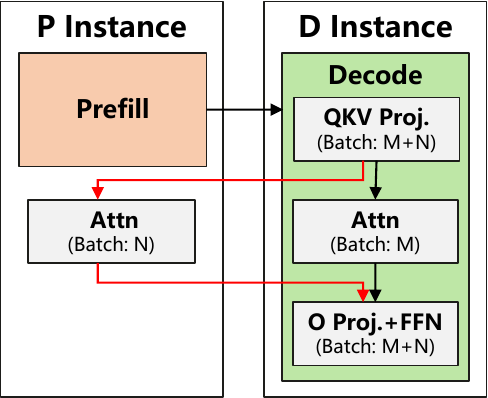}}
        \vspace{-0.8em}
	\caption{\label{fig:disaggArch} \needupdate{Comparison of original PD disaggregation and \ours{} (decoding batch size: $M$ v.s. $M$ + $N$).}}
\end{figure}

Placing both the prefill and decoding phases on the same GPUs enables
shared model parameters, but it can lead to performance
interference between the two phases.
The SLOs for the prefill and decoding phases differ significantly. The durable
latency for prefill (i.e., TTFT) is typically longer than that of the decoding phase
(i.e., TPOT)~\cite{DBLP:conf/osdi/ZhongLCHZL0024}. 
Executing the compute-intensive prefill phase alongside decoding iterations
can slow down the decoding phase. Moreover,
accurately forecasting the sequence length
and request rate in real-world LLM inference scenarios is challenging.
Static allocating of GPUs during the system
initialization may fail to satisfy the dynamic demands 
of both phases, resulting in SLO violations for TTFT and TPOT, or underutilized GPU resources due to over-provision.

Given the distinct characteristics of the prefill and decoding phases,
several PD disaggregation designs~\cite{DBLP:conf/isca/PatelCZSGMB24,
DBLP:conf/osdi/ZhongLCHZL0024,DBLP:journals/corr/abs-2407-00079} are proposed to assign the two phases
to separate GPUs. PD disaggregation effectively eliminates the performance interference of the two phases and enables flexible scaling of GPU resources for
each phase on demand. Figure~\ref{fig:disaggArch} shows the basic architecture of existing PD disaggregation systems. 
A prefill or decoding instance indicates a group of GPUs dedicated to
running either the prefill or decoding phase of LLM inference.
Once the prefill phase is completed, the corresponding KV caches
are transferred to the decoding instances via high-bandwidth
inter-connections such as NVLink and InfiniBand.

\subsection{Observation and Motivation}
\label{sec:motivation}

\begin{figure}[t!]
	\centering
        \hspace*{2em}\includegraphics[width=0.8\linewidth]{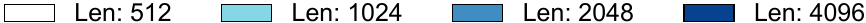} \\ [-0.6em]
	\subfloat[\label{fig:PKCompUtil} Compute utilization]{
		\includegraphics[width=0.48\linewidth]{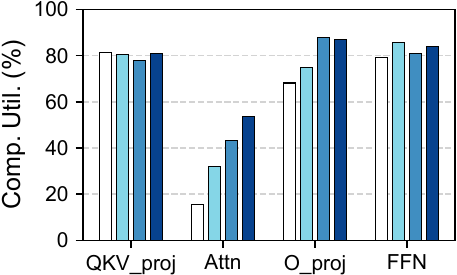}}\hfill
	\subfloat[\label{fig:PKBwUtil} HBM bandwidth utilization]{
		\includegraphics[width=0.48\linewidth]{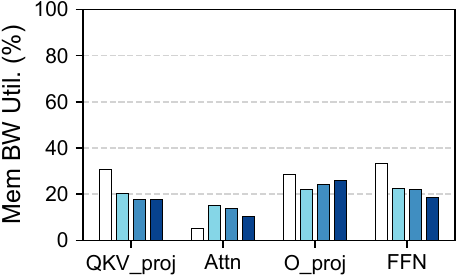}}
	\caption{\label{fig:PKUtil} \needupdate{Resource utilization of main kernels in prefill instances running Llama-2 7B with different prompt lengths (batch size: 1).}}
\end{figure}

\begin{figure}[t!]
	\centering
        \hspace*{2em}\includegraphics[width=0.8\linewidth]{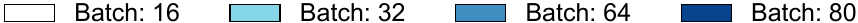} \\ [-0.6em]
        \subfloat[\label{fig:DKCompUtil} Compute utilization]{
		\includegraphics[width=0.48\linewidth]{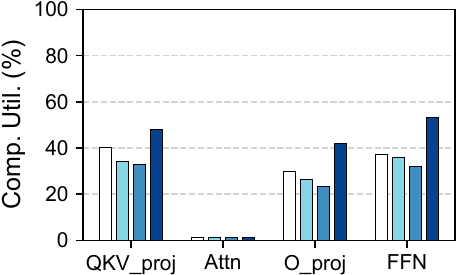}} \hfill
        \subfloat[\label{fig:DKBwUtil} HBM bandwidth utilization]{
		\includegraphics[width=0.48\linewidth]{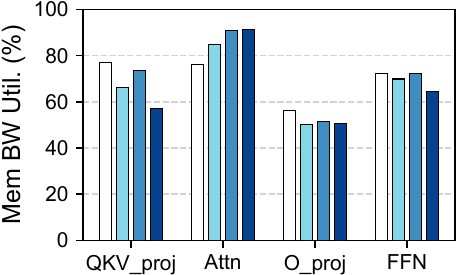}}
	\caption{\label{fig:DKUtil} \needupdate{Resource utilization of main kernels in decoding instances running Llama-2 7B with different batch sizes (sequence length: 1K).}}
    \vspace{-0.8em}
\end{figure}

Although PD disaggregation provides flexibility, we observe that it leads to intra-instance resource underutilization. As shown in Figure~\ref{fig:PDUtil}, both the HBM bandwidth utilization in prefill instances and the compute resource utilization in decoding instances are low. To further investigate, we measure the resource utilization of underlying kernels in both the prefill and decoding phases. As shown in Figures~\ref{fig:PKUtil} and \ref{fig:DKUtil}, four key kernels in the prefill phase — QKV projection, attention computation, output projection, and FFN — are compute-intensive compared to that of the decoding phase, making the memory bandwidth underutilized. Moreover, the KV cache of requests is immediately sent to decoding instances once the prefill phase is finished, causing low HBM capacity utilization in the prefill phase (Figure~\ref{fig:PDHbmUtil}).
Unlike the prefill phase, the auto-regression feature of the decoding phase results in the low arithmetic intensity of GPU kernels. As a result, compared to the prefill instance, the compute utilization rates of four decoding kernels are much lower, as shown in Figure~\ref{fig:DKUtil}. The decoding instance often suffers from limited HBM capacity (Figure~\ref{fig:PDHbmUtil}) and bandwidth (Figure~\ref{fig:DKBwUtil}). In the decoding phase, as shown in Figure~\ref{fig:AttnTimeRatio}, the attention computation kernel dominates the execution time with the increase of the decoding batch size. When the batch size is 80, the attention kernel accounts for 69.5\% execution time per transformer layer. In general, the attention computation kernel becomes the primary consumer of HBM capacity and bandwidth in the decoding phase. 

To enhance GPU resource utilization in both prefill and decoding instances, we propose disaggregating and offloading part of the memory-intensive attention tasks from the decoding phase to GPUs running the prefill phase. Executing the offloaded attention tasks improves the utilization of both capacity and HBM bandwidth in prefill instances. Additionally, this enables an increase in the total batch size in decoding phase, thereby boosting compute resource utilization in the decoding instance.


\subsection{Challenges}
\label{sec:challenge}

Nevertheless, there are three main challenges to achieve efficient offloading attention
in \ours{}.

First, offloading attention introduces extra synchronization steps within each transformer layer in the decoding phase. Specifically, as shown in Figure~\ref{fig:disaggOffloadArch}, to offload attention computation, decoding instances must send input parameters (i.e., $qkv$) to prefill instances. Once the offloaded attention computation is completed, the output (i.e., $attn\_out$) must be sent back to decoding instances for further processing. These extra steps—sending $qkv$, executing offloaded attention, and receiving the output—must be carefully overlapped with the local attention computation in the decoding instances. Each of these steps involves CPU intervention to issue corresponding kernel executions to the GPUs. Small synchronization overheads per transformer layer accumulate into high decoding costs. For example, if the synchronization overhead is 0.5 ms for each layer, the total time added to the TPOT of a llama-2 7B model with 32 layers can be as much as 16 ms. 

Second, offloaded attention tasks consume GPU resources of prefill instances, potentially causing performance interference between the prefill and offloaded attention tasks. As shown in Figure~\ref{fig:prefillUtil}, compute resource utilization in the prefill phase is already high. Although attention computation itself requires relatively low compute power, running attention and prefill tasks concurrently can cause interference. A possible solution is to partition a GPU into two virtual GPUs (vGPUs)~\cite{NVIDIAVirtualGPU}. However, vGPUs are shared at the virtual machine level and the predefined resource partition configurations in vGPUs are often inefficient for the dynamic and unpredictable inference workloads encountered in real-world scenarios. NVIDIA's multi-instance GPU (MIG)~\cite{NVIDIAMIG} technology offers hardware support for splitting memory and compute resources, such as streaming multiprocessors (SMs). However, the computing power split by MIG is proportional to the allocated HBM capacity, which greatly degrades the prefill performance. Besides, it also faces similar limitations to vGPUs. Overall, the colocation of attention and prefill tasks requires a more flexible GPU resource partitioning strategy.

Third, accurately determining the amount of attention to offload (e.g., $N$ in Figure~\ref{fig:disaggOffloadArch}) is difficult. The offloading ratio must balance both compute and memory resource utilization in prefill and decoding instances. A straightforward method involves offline profiling using existing workloads. However, the prompt length and the output length for online LLM inference are unknown in advance and usually not consistent with previous workloads. A static offloading ratio derived from offline profiling often fails to perform well for real-world scenarios with dynamic workloads.

\section{The \ours{} Design}
\label{sec:design}

\subsection{Overview}
\label{sec:designOverview}

\begin{figure}[t!]
  \centering
  \includegraphics[width=0.9\linewidth]{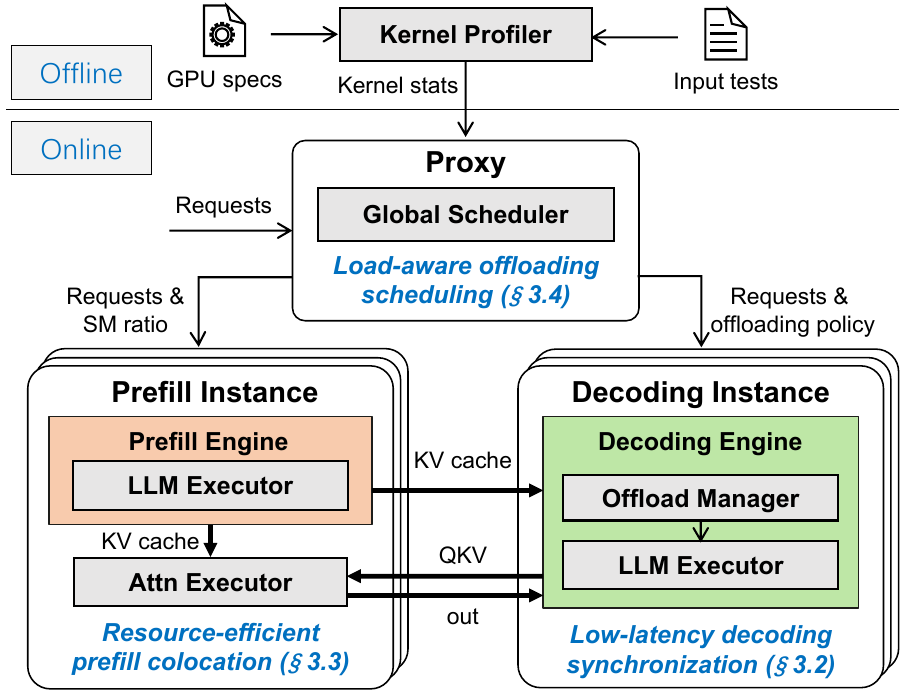}
  \caption{\label{fig:overview} \needupdate{The architecture overview
    of \ours{}}.}
\end{figure}

In this section, we present \textit{Adrenaline}, an attention disaggregation and offloading mechanism designed to enhance resource utilization and performance of LLM serving systems. The core idea of \textit{Adrenaline} is to disaggregate part of the attention computation in the decoding phase and offload them to the prefill instances. By leveraging the memory capacity and bandwidth of prefill instances, \textit{Adrenaline} improves their memory utilization, while simultaneously increasing the decoding batch size to enhance the compute utilization of decoding instances. 

The overview of \ours{} is shown in Figure~\ref{fig:overview}.
Similar to existing PD disaggregation designs~\cite{DBLP:conf/isca/PatelCZSGMB24,
DBLP:conf/osdi/ZhongLCHZL0024,DBLP:journals/corr/abs-2407-00079}, \ours{} consists of three main modules: \emph{proxy}, \emph{prefill instance}, and \emph{decoding instance}. The proxy module routes inference tasks to the appropriate prefill
and decode instances, where the respective engines execute
the corresponding prefill or decoding tasks during LLM inference.
However, unlike existing designs, \ours{} introduces the disaggregation and
offloading of a portion of the decoding attention computation to the remote \emph{attention executor},
which is colocated with the prefill engines. The attention executor is specifically dedicated to executing the
offloaded decoding attention tasks using the underutilized GPU memory
resources in the prefill instances.

To efficiently overcome the challenges outlined in \secref{sec:challenge}, \ours{} incorporates
three key optimizations in the LLM inference workflow.

First, in the decoding phase, \ours{} introduces a low-latency decoding synchronization mechanism
(\secref{sec:decodeOverlap}) to minimize the
synchronizing overhead between the remote attention executor
and the local decoding engine.

Second, by leveraging a resource-efficient prefill colocation 
design (\secref{sec:prefillColocation}), \ours{} enhances
the GPU memory utilization rates in prefill instances while ensuring sufficient compute resources for the
prefill phase. This eliminates performance interference, allowing the system to meet the required SLOs.

Third, \ours{} leverages a load-aware offloading scheduling strategy
(\secref{sec:proxyScheduling}) to adaptively determine
whether the attention computation of a request should be offloaded.  This strategy takes into account factors such as available GPU resources and current load.


\subsection{Low-latency Decoding Synchronization}
\label{sec:decodeOverlap}

\begin{figure}[t!]
  \centering
	\subfloat[\label{fig:overlapA} Original decoding phase without attention offloading]{
		\includegraphics[width=0.98\linewidth]{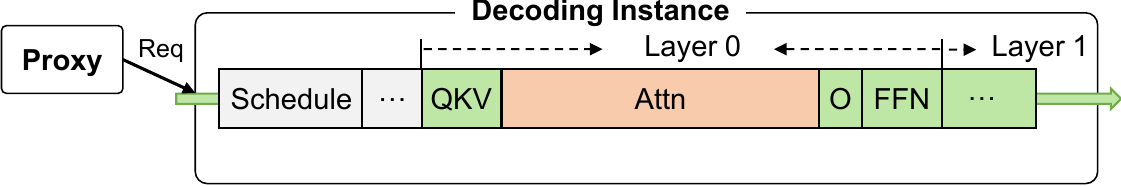}}
  \hfill
	\subfloat[\label{fig:overlapB} Decoding phase with low-latency attention offloading]{
		\includegraphics[width=0.98\linewidth]{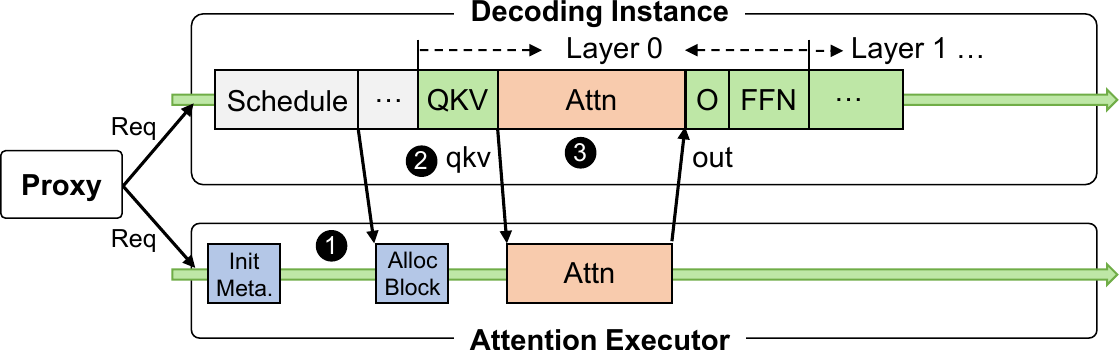}}
  \caption{\label{fig:overlap} \needupdate{Comparison of decoding the same batch size
    of requests without/with attention offloading in \ours{}}.}
\end{figure}

Figure~\ref{fig:overlap} compares the workflows of the decoding phase with and without attention offloading. In \ours{} (\ref{fig:overlapB}), the remote executor in the prefill instance executes the offloaded attention computation in parallel with the local attention computation in the decoding instance, improving the parallel throughput. However, to realize the performance benefits, it is critical
to minimize the operation scheduling and synchronization overheads between the local attention execution and the remote attention offloading.
If not carefully managed, the accumulated overheads per layer can lead to increased decoding latency, negating the potential improvements in throughput.
To address this problem, \ours{} achieves low-latency decoding synchronization by carefully optimizing the attention offloading workflow and kernel launching.



\subsubsection{Offloading Workflow Orchestration}
The critical path of the attention offloading workflow includes the operations started from sending $qkv$ and ended by receiving the attention output. If the critical path is longer than executing local attention in the decoding instance, the inference stalls due to waiting for the output of offloaded attention.

To shorten the critical path, \ours{} carefully organizes the attention offloading workflow. First, metadata and memory management operations for offloaded attention are moved outside the critical path (\circled{1} in Figure~\ref{fig:overlapB}). When the proxy determines that a request's attention needs to be offloaded (Algorithm~\ref{alg:NeedOffload}), the new request is sent as a hint to the attention executor. This allows the attention executor to initialize runtime metadata (e.g., prompt length) for the new request before entering the critical path. Similarly, resource management operations, such as allocating and recycling cache blocks, are performed outside the critical path. Second, instead of sending the scattered attention inputs (e.g., $q$, $k$, $v$) separately, \ours{} groups them and sends the aggregated data in a single operation (\circled{2} in Figure~\ref{fig:overlapB}). This reduces the communication costs for the input of the attention executor. Third, the scheduling algorithm in the proxy (\secref{sec:proxyScheduling}) ensures that the GPU time spent on the attention kernel in the attention executor is overlapped with the time spent on the local attention kernel in the decoding instance (\circled{3} in Figure~\ref{fig:overlapB}). This reduces idle waiting time and improves GPU resource utilization.

\subsubsection{Kernel Pre-launching}


Similar to existing PD disaggregation systems, attention offloading involves several small GPU kernels for each transformer layer, e.g., grouping $qkv$, sending $qkv$, receiving attention output, and merging outputs from two attention kernels. Though these small kernels execute quickly on GPUs, the CPU overhead for sequentially launching them can be non-negligible.
For example, when executing the decoding phase of Llama-2 7B using one A100 (batch size is 8 and the sequence length is 1K), the average GPU time per transformer layer is 0.38 ms but the average CPU time is 1.137 ms, thus wasting 0.76 ms GPU time per transformer layer for CPU overhead.
Instead of launching small kernels one by one, CUDA graph technology~\cite{CUDAGraph} captures multiple sequential kernels to be replayed as a single fused kernel, thereby efficiently improving the decoding performance (e.g, 2.6$\times$ under the aforementioned setting). However, the existing CUDA graph technique in vLLM becomes ineffective for attention offloading, since the unknown and dynamic numbers of locally executed requests and offloaded requests violate the requirement of static tensor shape in vLLM's CUDA graph, significantly complicating graph capture operations.

To mitigate the kernel launch overhead, \ours{} adopts a two-dimensional CUDA graph for attention offloading. The first dimension refers to the graph capacity for the original batch size in the decoding instance ($C_d$). The second dimension indicates the graph capacity for offloaded attention computation tasks ($C_o$). The combination of two parameters leads to $C_d\times C_o$ CUDA graphs, increasing the storage overhead. To address this problem, \ours{} sets configurable intervals to limit the total number of captured CUDA graphs.

Given an offloading ratio (\secref{sec:proxyScheduling}), one critical problem is the selection of a captured CUDA graph for each decoding iteration. By leveraging the scheduler output, the offloading manager in the decoding instance adopts the smallest two-dimensional CUDA graph that accommodates both local and remote attention batches. The selected two-dimensional CUDA graph is used by the paired decoding instance and attention executor. To mitigate synchronization costs, the result of the selected CUDA graph is sent to the attention executor outside the critical path of the model forward execution (\circled{1} in Figure~\ref{fig:overlapB}).




\subsection{Resource-efficient Prefill Colocation}
\label{sec:prefillColocation}

Offloaded attention is executed by the attention executor, which shares GPU resources with prefill instances. While this colocation improves GPU utilization, it also presents challenges, such as potential resource competition, which can lead to performance interference.

To address potential performance interference, we conduct a study on resource-constrained LLM inference performance. Based on the key observations from aforementioned study, we propose a resource-efficient colocation mechanism in the prefill instance to ensure performance isolation and improve the overall system performance.

\begin{figure}[t!]
   \begin{minipage}{0.48\linewidth}
     \centering
     \includegraphics[width=0.98\linewidth]{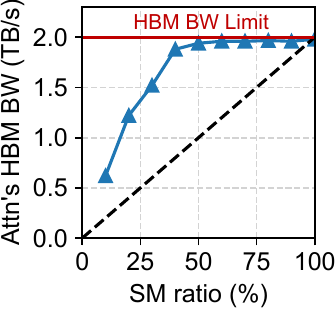}
     \caption{\label{fig:attnBW} \needupdate{The maximal memory bandwidth of the attention kernel using various ratios of SMs}.}
   \end{minipage}\hfill
   \begin{minipage}{0.48\linewidth}
     \centering
     \includegraphics[width=.98\linewidth]{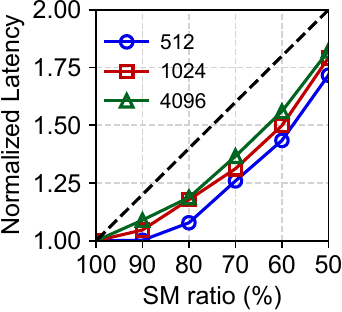}
     \caption{\label{fig:prefillSM} \needupdate{The normalized prefill throughput with different prompt lengths using various ratios of SMs}.}
   \end{minipage}
\end{figure}

\subsubsection{Resource Utilization Profiling}

To investigate the impact of resource partition for LLM inference, we evaluate the prefill phase and attention computation under various ratios of compute resources (SMs). Specifically, NVIDIA MPS technology~\cite{NVIDIAMPS} is adopted to limit the number of SMs for targeted computation tasks. In the study, we have obtained two observations as follows.

1) \emph{With the increase of used SMs, the HBM bandwidth utilization of the attention computation kernel super-linearly increases}. We measure the HBM bandwidth of the attention computation with different ratios of SMs using the Llama-2 7B model. As shown in Figure~\ref{fig:attnBW}, 
since the attention computation kernel is memory-intensive, even partial SMs are able to achieve high memory bandwidth. Due to the multi-level parallelism, e.g., SM-parallel, warp-parallel, and thread-parallel execution, in modern GPUs, the attention kernel demonstrates superior efficiency. For instance, 20\% SMs obtain 60\% of A100's HBM bandwidth.

2) \emph{With the decrease of used SMs, the prefill latency increases at a sub-linear level.} We constrain the SM cores and evaluate the prefill latency with the Llama-2 7B mode. \needupdate{Figure~\ref{fig:prefillSM} indicates that the prefill performance is closely related to the number of allocated SMs.} In PD disaggregation systems, some sub-steps in the prefill phase, e.g., request routing, scheduling, and KV cache transfer, do not rely on GPU compute resources. Hence, it is possible to limit the degradation of the prefill phase under partial SMs.

In summary, the key insight is non-linear relationship between the SM count and the performance of colocated components (i.e., attention bandwidth and prefill latency), which ensures the efficiency of resource sharing in the prefill instance.

\subsubsection{Adaptive Resource Partition}

Motivated by the above two observations, \ours{} leverages a resource-efficient scheme to colocate the prefill engine and the attention executor. The key idea is to partition SMs according to the TTFT SLO. There are two stages for performance isolation: \emph{offline profiling} and \emph{online serving}, which are detailed below.

In the offline profiling stage, the prefill latency of various prompt lengths is measured by the kernel profiler with different numbers of SMs. Since the prefill phase is compute-intensive, the execution time of a specific model is proportional to the involved FLOPs, which respond to the total length of all prompts.

In the online serving stage, \ours{} provides a configurable variable to set the ratio of allocated SMs. According to the TTFT SLO and the statistics obtained in the offline profiling stage, \ours{} gets the minimal SM ratio, which is used in NVIDIA MPS to constrain the compute resource for subsequent prefill phases.

\subsection{Load-aware Offloading Scheduling}
\label{sec:proxyScheduling}

For the scheduling in \ours, the key design goal is to achieve optimal
performance improvement while minimizing offloading overheads.
There are two main problems to be addressed in the offloading
strategy: 

\emph{(1) How many decoding attention tasks can
be offloaded?} A low offloading ratio underutilizes the idle
GPU memory resources in prefill instances, resulting in negligible overall performance improvement. Conversely, offloading too many decoding attention
tasks can cause generation stalls in the decoding phase
due to excessive synchronization overheads, hindering overall performance.

\emph{(2) Given an ideal offloading ratio, how to efficiently
determine whether offloading is necessary for a request in a dynamic workload?}
In real-world scenarios, the prompt and generated
output lengths are various across different user requests.
Moreover, the arrival rate of user requests can fluctuate.
Therefore, it is critical to design a low-overhead and fine-grained
scheduling mechanism that can adapt to these dynamic conditions and make efficient offloading decisions.


For the first problem, we formulate a model to obtain the upper bound of the offloading ratio (\secref{sec:OffloadingBound}). For the second problem, \ours{} leverages the runtime information in the proxy to enable balanced attention offloading (\secref{sec:LoadAwareness}) and fine-grained adaptive scheduling (\secref{sec:AdapativeScheduling}).

\subsubsection{Bounding Offloading Ratio}
\label{sec:OffloadingBound}

To address the first problem of determining an appropriate offloading ratio, our aim is to identify the bound that maximizes attention offloading benefits without compromising system efficiency. The core idea is to find how much the decoding batch size can be increased without negatively affecting the TPOT, since low latency is critical to the SLO attainments of PD disaggregation systems. Specifically, given $M$ decoding requests in the decoding instance as shown in Figure~\ref{fig:disaggOffloadArch}, the problem is to find the maximal value of additional requests $N$ without affecting TPOT. \ours{} considers the memory resources in the prefill instance for the memory-intensive attention kernel and the available compute resources in the decoding instance for non-attention kernels.

In terms of the attention kernel, the upper bound of the offloading ratio is constrained by the available HBM capacity and bandwidth in prefill instances to enable perfect overlapping for remote attention computation. Let the average number of prefill instances allocated for each decoding instance be denoted by $n$. $HBM_{pi}$ and $BW_{pi}$ denote the HBM capacity and bandwidth of prefill instance $i$ allocated for attention executors. $HBM_d$ and $BW_d$ indicate the corresponding memory resources in the decoding instance. In practice, $HBM_{pi}$, $HBM_d$, $BW_{pi}$, and $BW_d$ are obtained by the kernel profiler in the offline stage. The upper bound of the offloading ratio in terms of memory resources, denoted by $OB_{mem}(n)$, can be calculated as:
\begin{equation}\label{eq:OffloadRatioMem}
OB_{mem}(n)=min\bigg( \frac{\sum_{i=1}^{n}HBM_{pi}}{HBM_d}, \frac{\sum_{i=1}^{n}BW_{pi}}{BW_d} \bigg)
\end{equation}

For the non-attention kernels, increasing the decode batch size also consumes more compute resources in the decoding instance, which leads to a second upper bound on the offloading ratio. In the decoding phase, the arithmetic intensity (i.e., FLOP:I/O ratio) of non-attention kernels such as FFN and linear projection for $QKV$ and $O$ is given by: $O(1/(1/h+1/b))$, where $h$ denotes the hidden state size (a model-specific constant) and $b$ is the batch size. As the decoding batch size increases, the arithmetic intensity also increases. However, when the non-attention kernels remain memory-bound, the execution time for these non-attention kernels is stable across different batch sizes. The boundary of arithmetic intensity between memory-bound and compute-bound for non-attention kernels is determined by GPU hardware characteristics (e.g., FLOPs and memory bandwidth) and model parameters (e.g., $h$). To accurately identify this boundary, offline profiling is required to determine the batch size at which the execution time of non-attention kernels increases with increased batch size. Let $B_{max}$ denote the maximal batch size for memory-bound non-attention kernels. Given a TPOT SLO, $B_{TPOT}$ denotes the maximum batch size that a decoding instance can handle without offloading, which is collected in the proxy during runtime. Based on these measurements, the upper bound of the offloading ratio in terms of compute resources can be calculated as:
\begin{equation}\label{eq:OffloadRatioComp}
OB_{comp}(B_{max})=\frac{B_{max}-B_{TPOT}}{B_{TPOT}}
\end{equation}

Overall, considering both memory and compute resource constraints, the upper bound of the offloading ratio is given by the following equation:
\begin{equation}\label{eq:OffloadRatio}
OB(n,B_{max})=min\big( OB_{mem}(n), OB_{comp}(B_{max}) \big)
\end{equation}
$OB(n,B_{max})$ is the overall upper bound, which is determined by the minimum of $OB_{mem}(n)$ and $OB_{comp}(B_{max})$. The value of $OB_{mem}(n)$ is derived from offline profiling and is updated when the allocated resources from prefill instances change. On the other hand, $OB_{comp}(B_{max})$ is a dynamic threshold depending on factors such as the workload characteristics (e.g., sequence length and request rate) and the target TPOT SLO.

\subsubsection{Load Awareness} 
\label{sec:LoadAwareness}

A straightforward way to implement offloading strategies would be to manage them within the scheduler of each decoding instance. However, the local scheduler in a decoding instance lacks awareness of the dynamic scaling of prefill instances, which can expand or shrink on demand. 

\ours{} addresses this issue by employing a \emph{global scheduler} in the proxy, which manages the metadata of runtime information, called \emph{runtime metadata}, to efficiently support offloading in dynamic workloads. By routing both requests and responses, the proxy can easily track the number of active requests in the decoding phase along with their corresponding sequence lengths. With the runtime metadata, the proxy gains insight into the current load of decoding instances, enabling it to calculate $B_{TOPT}$ and, consequently, $OB_{comp}(B_{max})$ for the offloading ratio upper bound. The proxy is also aware of the available prefill instances, which allows it to update $OB_{mem}(n)$ whenever prefill instances are added or removed.


\subsubsection{Fine-grained Adaptive Scheduling}
\label{sec:AdapativeScheduling}

\begin{algorithm}[t!]
	\caption{Load-aware Offloading Scheduling Algorithm}
	\label{alg:NeedOffload}
\begin{algorithmic}[1]
 \Require a new request $req$, the upper bound of offloading ratio $OB(n,B_{max})$, local running requests in the decode instance $LR$, offloaded requests of the decode instance $OR$
 \Ensure flag $need\_offload$
    \State {$need\_offload \gets 0$}
    \State {$attn\_max\_tokens \gets \sum_{r \in OR} r.max\_token $}
    \State {$attn\_used\_tokens \gets \sum_{r \in OR} r.used\_token $}
    \State {$decode\_used\_tokens \gets \sum_{r \in LR} r.used\_token $}
    \If {$attn\_used\_tokens + req.max\_token<\newline
      \hspace*{1.5em} decode\_used\_tokens \times OB(n,B_{max})$ }  \Comment{C1} \label{lc:p2c1}
      \State {$need\_offload \gets 1$}
    \ElsIf {$(attn\_used\_tokens + req.used\_token < \newline
      \hspace*{1.5em} decode\_used\_tokens \times OB(n,B_{max})) \newline
      \hspace*{1.5em} \&\& (\left\lvert OR \right\rvert + 1 < \left\lvert LR \right\rvert \times OB(n,B_{max})) $} \Comment{C2} \label{lc:p2c2}
      \State {$need\_offload \gets 1$}
    \EndIf
    \State {$Return$ $need\_offload$}
	\end{algorithmic}
\end{algorithm}




To determine whether a request should be offloaded, \ours{} follows one principle in the adaptive scheduling strategy: offload the request only if the offloading ratio is within the corresponding upper bound $OB(n,B_{max})$. Offloading too many tasks to the remote attention executor would overload the prefill instances and increase the synchronization overhead in the decoding phase.

\ours{} leverages the runtime metadata in the proxy to adaptively schedule each request. The scheduling algorithm is outlined in Algorithm~\ref{alg:NeedOffload}. Following the proposed principle, attention offloading is enabled by setting $need\_offload = 1$ if it does not increase the decoding latency. Specifically, there are two alternative conditions under which offloading is permitted for the new request:


\begin{figure*}[t!]
	\centering
	\subfloat[\label{fig:SharegptTTFT} TTFT]{
		\includegraphics[width=0.24\textwidth]{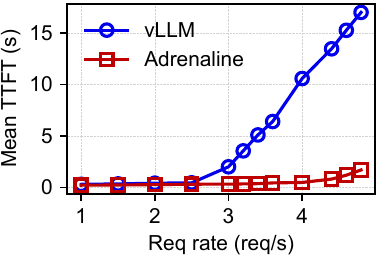}} \hfill
        \subfloat[\label{fig:SharegptTPOT} TPOT]{
      \includegraphics[width=0.24\textwidth]{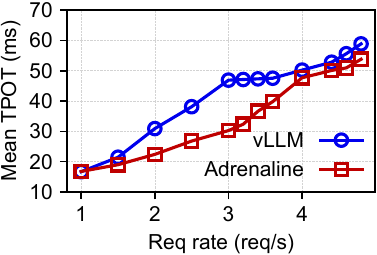}} \hfill
        \subfloat[\label{fig:SharegptP99TPOT} P99 TPOT]{
      \includegraphics[width=0.24\textwidth]{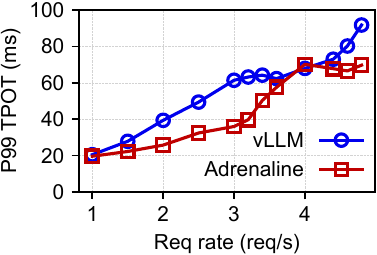}} \hfill
	\subfloat[\label{fig:SharegptThroughput} Output throughput]{
		\includegraphics[width=0.24\textwidth]{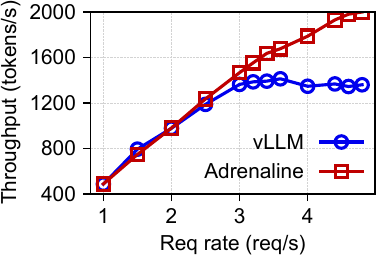}}
	\caption{\label{fig:E2ESharegpt} The E2E performance of ShareGPT with Llama-2 7B (\ours{}'s offloading ratio: 70\%).}
	\vspace{-0.2cm}
\end{figure*}

\begin{figure*}[t!]
	\centering
	\subfloat[\label{fig:SharegptLlama2-13BTTFT} TTFT]{
		\includegraphics[width=0.24\textwidth]{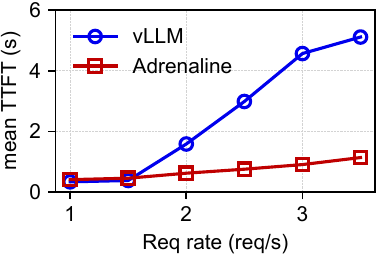}} \hfill
        \subfloat[\label{fig:SharegptLlama2-13BTPOT} TPOT]{
      \includegraphics[width=0.24\textwidth]{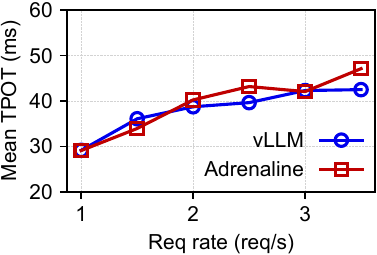}} \hfill
        \subfloat[\label{fig:SharegptLlama2-13BP99TPOT} P99 TPOT]{
      \includegraphics[width=0.24\textwidth]{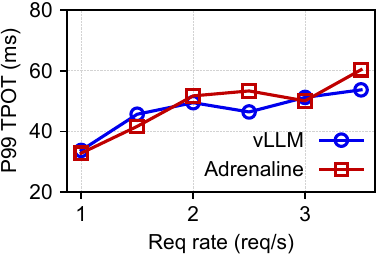}} \hfill
	\subfloat[\label{fig:SharegptLlama2-13BThroughput} Output throughput]{
		\includegraphics[width=0.24\textwidth]{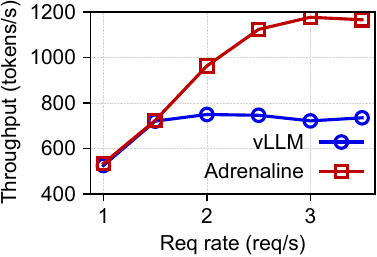}}
	\caption{\label{fig:E2ESharegptLlama2-13B} The E2E performance of ShareGPT with Llama-2 13B (\ours{}'s offloading ratio: 70\%).}
	\vspace{-0.2cm}
\end{figure*}

\begin{itemize}
    \item \textbf{C1}: Running attention of existing offloaded requests and the new request even with the maximal sequence length does not exceed the offloading ratio bound. This checking for sequence lengths ensures that the maximum of the computation time in the attention executor can be overlapped by the local attention run time, which enables low-latency synchronization for attention offloading.
    \item \textbf{C2}: For the attention executor and the decoding instance, both the ratios of current sequence lengths and the attention batch size are within the offloading ratio bound. Since the decoding instance and the corresponding attention executor are executed at the same frequency, checking attention batch sizes ensures that the load of the attention executor does not introduce decoding synchronization costs in the following iterations.
\end{itemize}

If C1 or C2 is satisfied, the new request's attention can be offloaded without increasing the decoding latency.

\section{Evaluation}
\label{sec:evaluation}

\subsection{Experimental Setup}
\label{sec:experimentalSetup}

\noindent
\textbf{Testbed.}
Our experiments are conducted on \needupdate{a high-performance server to evaluate the efficiency of \ours{} in LLM inference. The server contains eight NVIDIA A100-80GB SMX GPUs connected with 600 GB/s NVLink.}

\noindent
\textbf{Models.}
Our evaluation uses the popular Llama-2~\cite{DBLP:journals/corr/abs-2307-09288} 7B and 13B models with the default precision (float16).

\noindent
\textbf{Baseline.}
\ours{} is implemented based on vLLM~\cite{DBLP:conf/sosp/KwonLZ0ZY0ZS23}, which is a widely used open-source LLM inference framework. By default, we use vLLM as the baseline for performance comparison. The evaluated version (v6.3.0) of vLLM has already supported PD disaggregation.

\noindent
\textbf{Workloads.}
We evaluate performance using real-world workloads including ShareGPT~\cite{sharegpt} and OpenThoughts~\cite{openthoughts}. ShareGPT represents a chatbot scenario and has been adopted by related LLM inference research~\cite{DBLP:conf/osdi/ZhongLCHZL0024, DBLP:conf/usenix/GaoHSKJDYYZ24}. OpenThoughts~\cite{openthoughts} is a popular dataset of recent emerging reasoning models~\cite{DBLP:journals/corr/abs-2501-12948,DBLP:journals/corr/abs-2409-18486}. Both workloads consist of variable-length prompts, reflecting the dynamic nature of real-world LLM inference workloads. In general, due to the chain-of-thoughts~\cite{DBLP:conf/nips/Wei0SBIXCLZ22} at the beginning of the generation output, the output/prompt ratio in OpenThoughts is greater than that of ShareGPT.

\noindent
\textbf{Metrics.}
In the evaluation, TTFT and TPOT are respectively used to measure the prefill and decoding latency. Note that the TTFT in the PD disaggregation version of vLLM includes the time to transfer the KV cache from the prefill to the decoding instance. \ours{} adopts the same metric definition of TTFT following vLLM. In addition, we evaluate the token generation throughput in the decoding phase, referred to as \emph{output token throughput}. To avoid the impact of system warm-up stage, we record and report the output token throughput when the compared systems reach stable equilibrium states, which reflect the best generation throughput of different systems. Specifically, we take the duration between the first and last time when the HBM capacity of decoding instances is saturated (e.g., preemption in vLLM) as the stable state. If the request rate is too low to saturate the HBM space of decoding instances, we report the average output token throughput when the decoding batch size reaches 80\% of its peak batch size.

\subsection{End-to-End Performance}
\label{sec:e2ePerformance}


\begin{figure*}[t!]
	\centering
	\subfloat[\label{fig:OpenthoughtsTTFT} TTFT]{
		\includegraphics[width=0.25\textwidth]{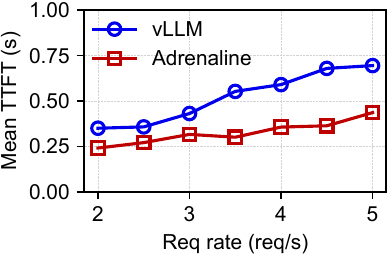}} \hfill
        \subfloat[\label{fig:OpenthoughtsTPOT} TPOT]{
      \includegraphics[width=0.24\textwidth]{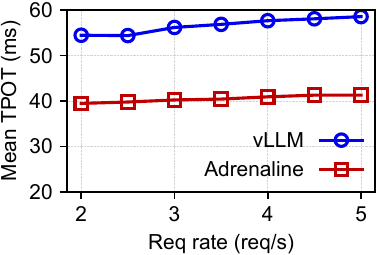}} \hfill
        \subfloat[\label{fig:OpenthoughtsP99TPOT} P99 TPOT]{
      \includegraphics[width=0.24\textwidth]{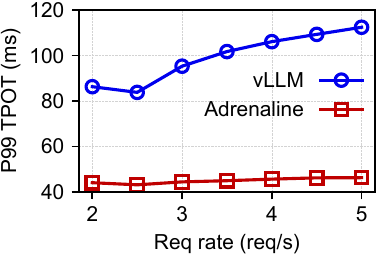}} \hfill
	\subfloat[\label{fig:OpenthoughtsThroughput} Output throughput]{
		\includegraphics[width=0.24\textwidth]{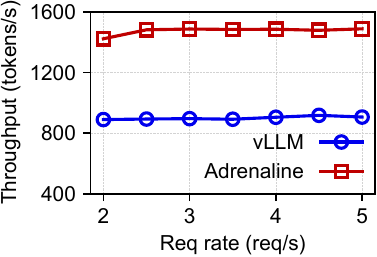}}
	\caption{\label{fig:E2EOpenthoughts} The E2E performance of OpenThoughts with Llama-2 7B (\ours{}'s offloading ratio: 80\%).}
	\vspace{-0.2cm}
\end{figure*}

\begin{figure*}[t!]
	\centering
	\subfloat[\label{fig:OpenthoughtsLlama2-13BTTFT} TTFT]{
		\includegraphics[width=0.25\textwidth]{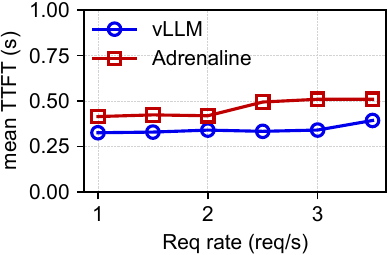}} \hfill
        \subfloat[\label{fig:OpenthoughtsLlama2-13BTPOT} TPOT]{
      \includegraphics[width=0.24\textwidth]{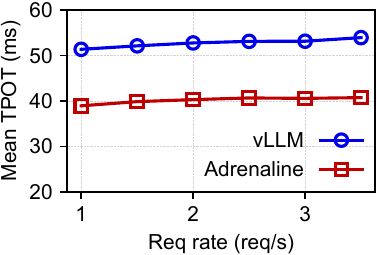}} \hfill
        \subfloat[\label{fig:OpenthoughtsLlama2-13BP99TPOT} P99 TPOT]{
      \includegraphics[width=0.24\textwidth]{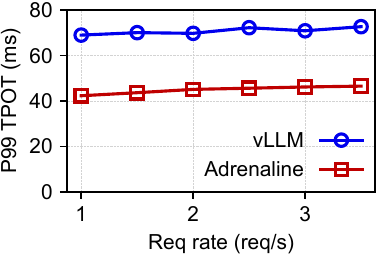}} \hfill
	\subfloat[\label{fig:OpenthoughtsLlama2-13BThroughput} Output throughput]{
		\includegraphics[width=0.24\textwidth]{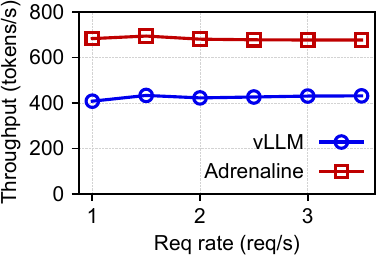}}
	\caption{\label{fig:E2EOpenthoughtsLlama2-13B} The E2E performance of OpenThoughts with Llama-2 13B (\ours{}'s offloading ratio: 80\%).}
	\vspace{-0.2cm}
\end{figure*}

\begin{figure*}[t!]
  \begin{minipage}{0.74\linewidth}
    \centering
    \hspace*{0.8em}\includegraphics[width=0.8\linewidth]{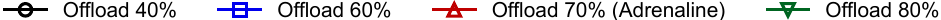} \\ [-0.6em]
    \subfloat[\label{fig:SharegptVaryRatioTTFT} TTFT]{
        \includegraphics[width=0.32\linewidth]{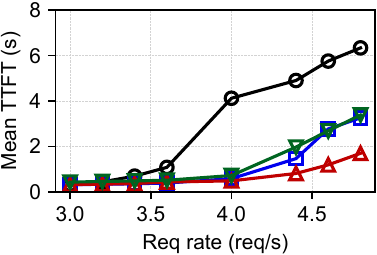}} \hfill
    \subfloat[\label{fig:SharegptVaryRatioTPOT} TPOT]{
        \includegraphics[width=0.32\linewidth]{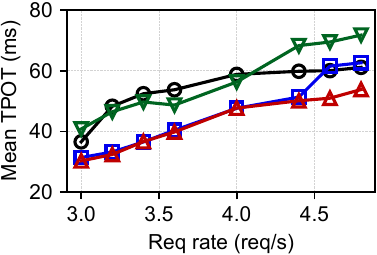}} \hfill
    \subfloat[\label{fig:SharegptVaryRatioTHPT} Output throughput]{
        \includegraphics[width=0.31\linewidth]{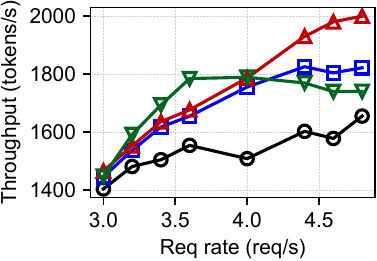}}
    \caption{\label{fig:SharegptVaryRatio} \needupdate{ShareGPT performance with Llama-2 7B using different offloading ratios.}}
  \end{minipage}\hfill
  \begin{minipage}{0.24\linewidth}
    \centering
     \includegraphics[width=.98\linewidth]{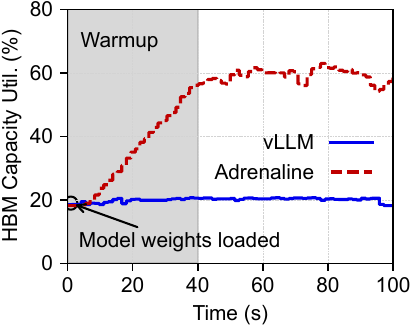}
    \caption{\label{fig:E2EHBMUtil} \needupdate{HBM use in the prefill instance}.}
  \end{minipage}
\end{figure*}

In this section, we evaluate the end-to-end performance of different schemes to demonstrate the efficiency of \ours{} under real-world workloads.



\textbf{TTFT.}
We measure the TTFT of both 7B and 13B models under various request rates, as shown in Figures~\ref{fig:E2ESharegpt}-\ref{fig:E2EOpenthoughtsLlama2-13B}. For ShareGPT in Figures~\ref{fig:SharegptTTFT} and \ref{fig:OpenthoughtsTTFT}, when the request rate is low, the TTFT of \ours{} is close to that of vLLM. However, by increasing the request rate of ShareGPT, the decoding instance of vLLM exhausts the HBM resource, thus blocking new decoding requests and significantly prolonging the queueing time. For example, when the request rate is 4 for ShareGPT with Llama-2 7B, the TTFT of vLLM is 22$\times$ higher than \ours{}.   By offloading some memory-intensive attention tasks, \ours{} increases the maximal batch size of the decoding phase. As a result, the queueing time in \ours{} is mitigated and the TTFT is reduced. Since the prompt lengths of requests in OpenThoughts are short, the TTFTs of both schemes are similar.







\textbf{TPOT.}
As shown in Figures~\ref{fig:SharegptTPOT}-\ref{fig:OpenthoughtsLlama2-13BTPOT}, the TPOT of \ours{} is often close to or lower than that of vLLM at the same request rate. In terms of TPOT, the benefits of \ours{} come from two factors. First, when the request rate is low (e.g., request rate < 3 in Figure~\ref{fig:SharegptTPOT}), \ours{} leverages the idle HBM bandwidth of prefill instances to execute offloaded attention computation tasks. The aggregated HBM bandwidth decreases the execution time of attention, which is the bottleneck of the decoding phase for large batch sizes (Figure~\ref{fig:AttnTimeRatio}). However, if the HBM space of the decoding instances is exhausted, the batch size of offloaded attention in \ours{} is bounded by the execution time of local attention computation tasks (\secref{sec:proxyScheduling}). Hence, the benefits for TPOT diminish when the request rate is higher than 3 in Figure~\ref{fig:SharegptTPOT} or higher than 1.5 in Figure~\ref{fig:OpenthoughtsTPOT}. Second, \ours{} mitigates the impact of preemption (or swapping), which happens when the HBM capacity of the decoding instance is insufficient to accommodate generated outputs of ongoing requests. A preempted request cannot be scheduled until the completion of at least one previous request to release HBM space. The preemption overhead is high for workloads with long output sequences, e.g., OpenThoughts. As shown in Figures~\ref{fig:OpenthoughtsTPOT} and \ref{fig:OpenthoughtsLlama2-13BTPOT}, compared to vLLM, \ours{} mitigates the preemption overhead, and hence reduces the mean TPOT by 26.9\%-29.5\% and 23.4\%-24.5\% for 7B and 13B models, respectively, with OpenThoughts.



The 99th percentile (P99) TPOT is also evaluated using real-world workloads. Similar to the mean TPOT, the low-latency decoding synchronization ensures that the P99 TPOT is close to that of vLLM, as shown in Figures~\ref{fig:SharegptP99TPOT} and \ref{fig:OpenthoughtsP99TPOT}. For OpenThoughts, due to the aggregated HBM bandwidth and the mitigation of decoding preemption, the P99 TPOT is reduced by 48.5\%-58.8\% and 34.9\%-38.7\% for 7B and 13B models, respectively, compared to vLLM.

\textbf{Output Token Throughput.}
We evaluate the output token throughput using various request rates. As shown in Figures~\ref{fig:SharegptThroughput}-\ref{fig:OpenthoughtsLlama2-13BThroughput}, both schemes achieve similar throughputs at low request rates. The throughput of vLLM reaches a plateau due to the limited HBM space and bandwidth in the decoding instance. \ours{} improves the throughput by leveraging the HBM resource of prefill instances, thus achieving up to 1.47$\times$ speedup for Llama-2 7B (1.63$\times$ for Llama-2 13B) with ShareGPT. For OpenThoughts, \ours{} offloads more attention (80\%) and obtains 1.60$\times$-1.66$\times$ speedup for Llama-2 7B (1.57$\times$-1.68$\times$ for Llama-2 13B).


\subsection{Offloading Ratio Selection}

Figure~\ref{fig:SharegptVaryRatio} shows the E2E performance of ShareGPT using different offloading ratios. When the offloading ratio is low, \ours{} improves the inference performance with the increase of the offloading ratio. As shown in Figure~\ref{fig:SharegptVaryRatio}, the performance of offloading 80\% attention is lower than the offloading ratio of 70\% due to the long run time of the attention executor, which cannot be overlapped by the local attention of the decoding instance. \ours{} obtains the inflection point for attention offloading according to the kernel statistics from the offline profiling and the workload patterns collected from the runtime metadata, thus automatically achieving high performance for different workloads.

\subsection{Resource Utilization}
\label{sec:testUtilization}


\begin{figure}[t!]
	\centering
        \hspace*{2em}\includegraphics[width=0.9\linewidth]{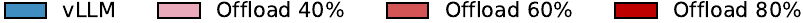} \\ [-0.4em]
        \subfloat[\label{fig:E2EBwUtil} HBM bandwidth utilization in the prefill instance]{
		\includegraphics[width=0.48\linewidth]{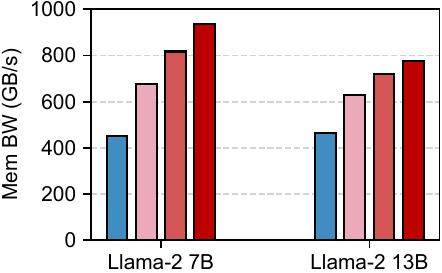}} \hfill
        \subfloat[\label{fig:E2ECompUtil} Compute utilization in the decoding instance]{
		\includegraphics[width=0.46\linewidth]{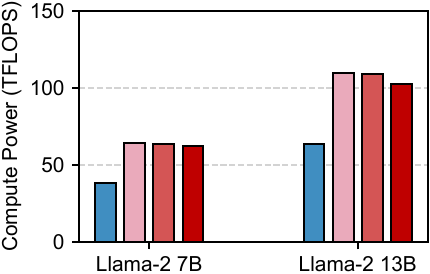}}
	\caption{\label{fig:E2EGpuUitl} \needupdate{Resource utilization with different offloading ratios.}}
\end{figure}

\begin{figure}[t!]
	\centering
        \hspace*{2em}\includegraphics[width=0.9\linewidth]{figures/decode_ops_compute_power_with_legend_cropped.pdf} \\ [-0.4em]
        \subfloat[\label{fig:PKBwUtil} HBM bandwidth in the prefill instance]{
		\includegraphics[width=0.36\linewidth]{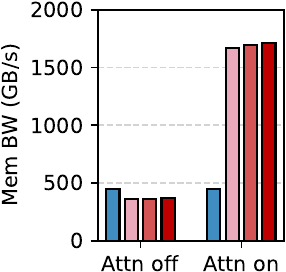}} \hfill
        \subfloat[\label{fig:DKCompUtil} Compute power of different kernels in the decoding instance]{
		\includegraphics[width=0.54\linewidth]{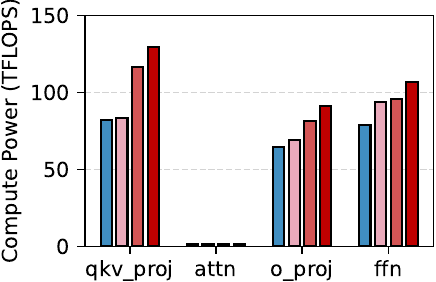}}
	\caption{\label{fig:UtilBreakdown} \needupdate{Resource utilization breakdown. (\texttt{Attn on} indicates the period when running offloaded attention)}}
\end{figure}

In this section, to demonstrate the efficiency of \ours{} for the imbalanced GPU resource utilization problem, we evaluate the utilization of HBM capacity and bandwidth in the prefill instance and the compute resource in the decoding instance. The evaluation of HBM capacity utilization uses the Llama-2 7B model and others are conducted with different models. ShareGPT workload is used in all experiments.

\textbf{HBM Capacity Utilization.}
As shown in Figure~\ref{fig:E2EHBMUtil}, after loading model weights, the HBM capacity utilization of the prefill instance in vLLM is around 20\%, leaving the remaining HBM capacity unused. Attention executor of \ours{} leverages the unused HBM capacity of the prefill instance to store KV caches for offloaded attention. As a result, Figure~\ref{fig:E2EHBMUtil} shows \ours{} achieves 2.28$\times$ HBM utilization ratio in the prefill instance after the warmup stage compared to the default PD disaggregation scheme.

\textbf{HBM Bandwidth Utilization.}
Figure~\ref{fig:E2EBwUtil} shows that offloading attention improves the HBM bandwidth of the prefill instance, since the attention executor computes aggregated memory-bound attention tasks. Specifically, compared to the baseline, \ours{} achieves 1.49$\times$-2.07$\times$ HBM bandwidth utilization for Llama-2 7B in the prefill instance. The trend for Llama-2 13B is similar (1.37$\times$-1.93$\times$ compared to vLLM). To investigate the HBM bandwidth consumption, we profile the memory bandwidth of the prefill instance when the attention executor is running (denoted by \texttt{Attn on}) and idle (denoted by \texttt{Attn off}). As shown in Figure~\ref{fig:PKBwUtil}, the attention executor obtains 3.76$\times$ HBM bandwidth (83.0\% of bandwidth capacity limit) on average than the prefill instance in the default PD disaggregation scheme. Note that ShareGPT only involves the attention executor for 23.0\%-42.9\% of run time. Workloads offloading more KV cache would increase the run time ratio of attention executor and hence consume more HBM bandwidth in the prefill instance.

\textbf{Compute Power Utilization.}
As shown in Figure~\ref{fig:E2ECompUtil}, by disaggregating attention computation, \ours{} increases the decoding batch size and hence achieves 1.67$\times$ compute power for Llama-2 7B (1.68$\times$ for Llama-2 13B) compared to vLLM. Note that the allocated CUDA cores are still able to accommodate the increased batch size when the offloading ratio is between 40\% and 80\%, causing the 
unchanged compute power for \ours{} in Figure~\ref{fig:E2ECompUtil}. Inside the decoding phase, the compute loads of the main GPU kernels are proportional to the batch size. Figure~\ref{fig:DKCompUtil} shows the compute power breakdown of four main kernels. With the increase of offloading ratio, the compute loads of three non-attention GPU kernels (i.e., QKV projection, O projection, and FFN) also increase. Specifically, \ours{} improves 8.8\% compute power in three non-attention kernels on average when offloading 40\% attention. If the offloading ratio is 80\%, the average improvement becomes 44.7\% compared to vLLM.

\section{Related Works}
\label{sec:relatedwork}

\noindent
\textbf{Disaggregated LLM Serving.} Recent efforts have advocated prefill and decoding disaggregation for efficient LLM serving. DistServe \cite{DBLP:conf/osdi/ZhongLCHZL0024} assigns prefill and decoding computation to separate GPUs to reduce interference, and configures resources for each phase independently. Splitwise \cite{patel2024splitwise} optimizes the cost, throughput, and power consumption of disaggregated LLM serving systems by jointly considering these factors.  Mooncake~\cite{qin2024mooncake} leverages the CPU, DRAM, and SSD resources to implement a disaggregated KV cache pool in disaggregated LLM serving systems. Our paper reveals the low resource utilization problem in disaggregated LLM serving systems and proposes \ours{} to address the problem through attention disaggregation and offloading.

\noindent
\textbf{Attention Offloading.} Several recent works have explored the potential of attention offloading to improve the performance of heterogeneous LLM serving systems. For example, Lamina~\cite{chen2024efficient} offloads the attention operation from high-end accelerators to a pool of lower-cost memory devices such as consumer-grade GPUs optimized for attention computation. Some designs~\cite{DBLP:journals/corr/abs-2403-11421,jiang2024neo} propose to offload the decoding attention computation to the CPU. NEO~\cite{jiang2024neo} offloads part of attention computation and KV cache states from the GPUs to the CPU machines, thereby increasing the GPU batch size and enhancing inference throughput. InstInfer~\cite{pan2024instinfer} utilizes computational storage devices (CSDs) to design a dedicated in-storage attention computation engine, reducing the overhead associated with transferring KV caches. Unlike these approaches, which rely on additional hardware resources for attention offloading, our proposed \ours{} improves inference performance by leveraging the underutilized resources within LLM serving systems, enabling efficient attention offloading without the need for extra hardware.

\section{Conclusion}
This paper addresses GPU resource underutilization in LLM serving systems with PD disaggregation. We propose \ours{}, an attention disaggregation and offloading mechanism that enhances memory and compute utilization by offloading decoding attention tasks to prefill instances. Key techniques include low-latency synchronization, resource-efficient prefill colocation, and load-aware offloading scheduling. Experimental results demonstrate that \ours{} significantly improves GPU utilization, increases throughput, and meets service-level objectives, providing a scalable and cost-effective solution for optimizing LLM inference performance.

\bibliographystyle{ACM-Reference-Format}
\bibliography{main}

\end{document}